# IRREVERSIBILITY IN CLASSICAL KINETIC THEORY: RETARDATION OF INTERACTION AND DISTRIBUTION FUNCTIONS

V.V.Zubkov[1,2], A.V.Zubkova[3]

[1]Department of General Physics, Tver State University, Tver, 170002, Russia

[2] Department of General and Experimental Physics, Yaroslav-the-Wise Novgorod State University, Veliky Novgorod, 173003, Russia

[3]Department of General Physics, Tver State Technical University, Tver, 170026, Russia

[1]*e-mail: victor.v.zubkov@gmail.com*

**Abstract**. A kinetic equation is derived for the phase density of a system of point particles, generating a system of integro-differential equations for distribution functions that have a deterministic meaning. The derivation took into account the retardation of interactions between particles. The obtained equations describe the irreversible behavior of a system of particles without involving any probabilistic hypotheses. The use of the retarded potentials, in this case, corresponds to taking the field into account when the dynamics of the many-particle system is described.

**Highlights**

- A new method for determining of distribution functions is proposed
- Distribution functions in the kinetic theory have deterministic nature
- Retarded interactions are considered as one of physical causes of irreversibility of the kinetic equations

**1. Introduction**

The main method for studying of nonequilibrium systems of many particles is the hierarchy of BBGKY kinetic equations for statistical distribution functions [1-3]. Formally, the BBGKY hierarchy can be obtained without involving distribution functions that have probabilistic meaning. For example, in [4] it was shown that, in the general case, it is possible to construct a chain of equations for microscopic phase densities similar to the BBGKY hierarchy, but without using of statistical assumptions and hypotheses.

Nevertheless, probabilistic hypotheses, ergodicity, and mixing formed the basis for explaining the transition of the system of many particles to equilibrium. However, the use of probabilities does not explain *from the physical point of view* neither the mechanism of transition to thermodynamic equilibrium nor the nature of thermodynamic equilibrium in general. Despite this, averaging with the use of probability measures is often used to obtain kinetic equations. For example, the Klimontovich equation [5], written for phase microscopic density, being completely deterministic and reversible in time, after formal statistical averaging goes into the first equation of the BBGKY hierarchy for statistical distribution functions. But such averaging does not contain the phenomenon of stochastization as a physical process and *does not explain* the transition to equilibrium state, but *only describes it*. It was also shown by Bogolyubov [6] that the stochasticization of the system is not contained even in the kinetic equations written for such a simple model as a system of elastic balls.

As a rule, a description of the transition of a system to equilibrium state is carried out either by averaging the solutions of the Liouville equation over the initial points in time in the distant past, or by adding an infinitely small source that selects delayed solutions of this equation [7]. These two methods are equivalent and are a generalization of the condition for the weakening of correlations according to Bogolyubov, i.e. actually contain an ergodic hypothesis [7]. But in the general case, the ergodic hypothesis is incorrect, and the criteria for the equality of time and configuration averages are unknown. In this regard, the properties of transitivity, and especially mixing for dynamical systems considered in statistical mechanics, are extremely difficult to establish [8]. In quantum statistics, in isolated dynamical systems, there is no analog of mixing at all, as well as the transitivity property, which, according to ergodic theory, is necessary for the transition of the system to equilibrium state [8].

We should also mention Katz's work [9], devoted to the problem of a discrete dynamic ring model. Katz showed that introducing realistic probabilistic hypotheses into a deterministic exactly solvable model leads to a distortion of the exact solution. Therefore, the justification of thermodynamics through the use of the concept of probability seems doubtful.

The mechanism of transition of the system to equilibrium state is often associated with the influence of thermostat [8]. Since a thermostat refers to the external environment, such a mechanism does not solve the problem of the transition to stochastization of an isolated system. This means that the reason for the transition to equilibrium state should be connected with interactions between particles of the system itself. However, in addition to the particles themselves, it is also necessary to take into account the field through which the particles interact.

The influence of the field leads to retarded potentials. Thus, one of the possible physical mechanisms generating irreversibility is not just the interaction of particles, but namely retarded interactions. Taking into account the retardation allowed Zakharov [10] to derive an irreversible equation for microscopic phase density, generalizing the Klimontovich's equation to the case of retarded interactions.

Finally, the mechanism of irreversible behavior of systems even with a small number of degrees of freedom was clarified in [11] and is also associated with the retardation of interaction between particles of the system.

In the present work, the *dynamic nature* of the distribution functions inherent in the kinetic equations of statistical physics is shown, and based on the consideration of the retarded interactions, the *kinetic equations irreversible in time* are obtained for them.

## 2. Determination of smoothed distribution functions

The description of the system of many particles at the microscopic level can be carried out using microscopic phase density [5]

$$f_{micro}(\mathbf{r},\mathbf{p},t) = \sum_{i=1}^{N} \delta(\mathbf{r}-\mathbf{r}_i(t))\delta(\mathbf{p}-\mathbf{p}_i(t)). \tag{1}$$

To proceed to the description of macroscopic properties, we adopt the following smoothing procedure [12]. We mentally divide the region occupied by a system of $N$ particles into small, time-independent volumes $\Delta(\mathbf{r})$. Let $N(\mathbf{r},t)$ is the number of particles in the volume $\Delta(\mathbf{r})$. The volume value $\Delta$ can be chosen quite arbitrarily depending on the research task. For example, if the number of particles of radius $r_0$ in such a volume is defined as $N(\mathbf{r},t) \approx (nr_0^3)^{-5/4}$ (here $n$ is the average concentration of particles), then using such a partition, a single description of both kinetic and gas-dynamic processes is possible [5].

Further, for each microscopic dynamic additive quantity $\chi_i$ to which the microscopic density $\chi(\mathbf{r},t) = \sum_{i=1}^{N} \chi_i \delta(\mathbf{r}-\mathbf{r}_i(t))$ corresponds, the field function $\psi(\mathbf{r},t)$ can be associated with the smoothing operator $\hat{S}$ [12]:

$$\psi(\mathbf{r},t) = \hat{S}[\chi(\mathbf{r},t)] \equiv \frac{1}{\Delta}\sum_{i=1}^{N(\mathbf{r},t)} \chi_i = \frac{1}{\Delta}\int_{\Delta} \chi(\mathbf{r}+\boldsymbol{\xi},t)d^3\boldsymbol{\xi}. \tag{2}$$

For example, the smoothed (or average) particle density in the volume $\Delta(\mathbf{r})$ has the form:

$$n(\mathbf{r},t) = \frac{1}{\Delta}\sum_{i=1}^{N(\mathbf{r},t)} m_i = \frac{1}{\Delta}\int_\Delta \sum_{i=1}^{N} m_i \delta(\mathbf{r}+\boldsymbol{\xi}-\mathbf{r}_i(t))d^3\xi. \qquad (3)$$

It should be emphasized that the operation (2) is a transition to the description of the dynamics in new (field) variables. Under the action of the operator $\hat{S}$, of course, most of the information about the motion of the particles of the system is lost, which, however, *does not lead to its irreversible behavior*. In statistical physics, namely the averaging (over any Gibbs ensemble) is considered as the origin of irreversibility. Here the situation is different. Just as the transition to the equation of motion of the center of inertia of a particle system in classical mechanics does not lead to irreversible behavior, so in our case operation (2) does not change the symmetry of the equation of motion with respect to time inversion.

We now define the macroscopic phase density as the smoothed microscopic density (1):

$$f(\mathbf{r},\mathbf{p},t) = \frac{1}{\Delta}\sum_{k=1}^{N}\int_\Delta d^3\xi\, \delta(\mathbf{r}+\boldsymbol{\xi}-\mathbf{r}_k(t))\delta(\mathbf{p}-\mathbf{p}_k(t)). \qquad (4)$$

It is significant that smoothing is carried out *only by the configuration subspace*. Carrying out additional smoothing over the momentum subspace, we would cut off some of the possible values of momenta of particles in the volume $\Delta$. This would lead, firstly, to inevitable errors, and secondly, could lead to *non-physical* non-invariance of the obtained equations of motion with respect to time reversal.

Integrating over the momentum subspace, we obtain a one-particle distribution function

$$F_1(\mathbf{r},t) = \int f(\mathbf{r},\mathbf{p},t)d^3\mathbf{p} = \frac{1}{\Delta}\int_\Delta \sum_{i=1}^{N}\delta(\mathbf{r}+\boldsymbol{\xi}-\mathbf{r}_i(t))d^3\xi,$$

coinciding with the particle number density (3). Similarly, we can introduce the two-particle distribution function, which will be used later:

$$\begin{aligned}f_2(\mathbf{r}_1,\mathbf{p}_1,\mathbf{r}_2,\mathbf{p}_2,t) = \frac{1}{\Delta}\int_\Delta \sum_{i=1}^{N}\sum_{k=1,k\neq i}^{N}\delta(\mathbf{r}_1+\boldsymbol{\xi}-\mathbf{r}_k(t))\delta(\mathbf{r}_2+\boldsymbol{\xi}-\mathbf{r}_i(t))\times\\ \times\delta(\mathbf{p}_1-\mathbf{p}_i(t))\delta(\mathbf{p}_2-\mathbf{p}_k(t))d^3\xi.\end{aligned} \qquad (5)$$

The above functions (4) and (5) are examples of dynamic distribution functions.

### 3. The basic kinetic equation

We find the partial time derivative of (4):

$$\frac{\partial f(\mathbf{r},\mathbf{p},t)}{\partial t} = -\frac{\mathbf{p}}{m}\frac{\partial f(\mathbf{r},\mathbf{p},t)}{\partial \mathbf{r}} - \sum_{k=1}^{N}\dot{\mathbf{p}}_k \frac{\partial}{\partial \mathbf{p}}\frac{1}{\Delta}\int_\Delta d^3\xi\, \delta(\mathbf{r}+\boldsymbol{\xi}-\mathbf{r}_k(t))\delta(\mathbf{p}-\mathbf{p}_k(t)). \qquad (6)$$

For further analysis, it is necessary to use the equation of motion of particles

$$\dot{\mathbf{p}}_i = -\frac{\partial}{\partial \mathbf{r}_i} U(\mathbf{r}_i, t). \tag{7}$$

The potential at the point $\mathbf{r}_i$, taking into account the retarded interactions, has the form:

$$U(\mathbf{r}_i, t) = \sum_{k \neq i}^{N} U(\mathbf{r}_i(t) - \mathbf{r}_k(t - \tau_{ik}))$$
$$= \sum_{k \neq i}^{N} \int U(\mathbf{r}_i(t) - \mathbf{r}_k(t')) \delta\left(t - t' - \frac{|\mathbf{r}_i - \mathbf{r}_k|}{c}\right) dt', \tag{8}$$

where $\tau_{ik} = \frac{|\mathbf{r}_i - \mathbf{r}_k|}{c} \equiv \tau(\mathbf{r}_i - \mathbf{r}_k)$ is the time of interactions retardation between $i$-th and $k$-th particles, $c$ is interaction transfer rate (the light speed). It should be specially noted that the function $U(\mathbf{r})$ is defined for resting particles, while the function (8) is determined not only by $\mathbf{r}$ and $t$, but also by retardation time $\tau_{ik}$ between all the particles.

Substituting (7) and (8) in (6), we rewrite the second term in (6) in the form

$$I = \sum_{i=1}^{N} \dot{\mathbf{p}}_i \frac{\partial}{\partial \mathbf{p}} \frac{1}{\Delta} \int_{\Delta(\mathbf{r})} d^3\xi \delta(\mathbf{r} + \xi - \mathbf{r}_i(t)) \delta(\mathbf{p} - \mathbf{p}_i(t)) =$$
$$= -\sum_{i=1}^{N} \sum_{k \neq i}^{N} \frac{1}{\Delta^2} \int_{\Delta(\mathbf{r}')} d^3\xi' \int_{\Delta(\mathbf{r})} d^3\xi \int d^3\mathbf{r}' \int d^3\mathbf{p}' \int dt' \frac{\partial G(\mathbf{r} + \xi - \mathbf{r}' - \xi', t, t')}{\partial \mathbf{r}} \times \tag{9}$$
$$\times \frac{\partial}{\partial \mathbf{p}} \delta(\mathbf{r} + \xi - \mathbf{r}_i(t)) \delta(\mathbf{p} - \mathbf{p}_i(t)) \delta(\mathbf{r}' + \xi' - \mathbf{r}_k(t')) \delta(\mathbf{p}' - \mathbf{p}_k(t')).$$

Here $G(\mathbf{r} + \xi - \mathbf{r}' - \xi', t, t') = \frac{\partial}{\partial \mathbf{r}} \left[ U(\mathbf{r} + \xi - \mathbf{r}' - \xi') \delta\left(t - t' - \frac{|\mathbf{r} + \xi - \mathbf{r}' - \xi'|}{c}\right) \right]$. After calculating the derivative with respect to $\mathbf{r}$ and integrating in (9) with respect to time $t'$ taking into account the ratio

$$\frac{\partial}{\partial t'} \delta(\mathbf{r}' + \xi' - \mathbf{r}_k(t')) = -\frac{\mathbf{p}_k(t')}{m} \frac{\partial}{\partial \mathbf{r}'} \delta(\mathbf{r}' + \xi' - \mathbf{r}_k(t')),$$

we rewrite (9) in a slightly different form

$$I = -\sum_{i=1}^{N} \sum_{k \neq i}^{N} \frac{1}{\Delta^2} \int_{\Delta(\mathbf{r}')} d^3\xi' \int_{\Delta(\mathbf{r})} d^3\xi \int d^3\mathbf{r}' \int d^3\mathbf{p}' G_1(\mathbf{r} + \xi - \mathbf{r}' - \xi') \times$$
$$\times \frac{\partial}{\partial \mathbf{p}} \delta(\mathbf{r} + \xi - \mathbf{r}_i(t)) \delta(\mathbf{p} - \mathbf{p}_i(t)) \delta(\mathbf{r}' + \xi' - \mathbf{r}_k(t - \tilde{\tau})) \delta(\mathbf{p}' - \mathbf{p}_k(t - \tilde{\tau})) -$$
$$- \frac{1}{\Delta^2} \int_{\Delta(\mathbf{r}')} d^3\xi' \int_{\Delta(\mathbf{r})} d^3\xi \int d^3\mathbf{r}' \int d^3\mathbf{p}' G_2(\mathbf{r} + \xi - \mathbf{r}' - \xi') \times$$
$$\times \frac{\mathbf{p}'}{m} \frac{\partial}{\partial \mathbf{p}} \sum_{i=1}^{N} \sum_{k \neq i}^{N} \frac{\partial}{\partial \mathbf{r}'} \delta(\mathbf{r} + \xi - \mathbf{r}_i(t)) \delta(\mathbf{p} - \mathbf{p}_i(t)) \delta(\mathbf{r}' + \xi' - \mathbf{r}_k(t - \tilde{\tau})) \delta(\mathbf{p}' - \mathbf{p}_k(t - \tilde{\tau})).$$

Here $G_1 = \dfrac{\partial U(\mathbf{r}+\boldsymbol{\xi}-\mathbf{r}'-\boldsymbol{\xi}')}{\partial \mathbf{r}}$, $G_2 = U(\mathbf{r}+\boldsymbol{\xi}-\mathbf{r}'-\boldsymbol{\xi}')\dfrac{1}{c}\dfrac{\mathbf{r}+\boldsymbol{\xi}-\mathbf{r}'-\boldsymbol{\xi}'}{|\mathbf{r}+\boldsymbol{\xi}-\mathbf{r}'-\boldsymbol{\xi}'|}$, and $\tilde{\tau}$ is the time of interactions retardation between points with coordinates $\mathbf{r}+\boldsymbol{\xi}$ and $\mathbf{r}'+\boldsymbol{\xi}'$: $\tilde{\tau} = |\mathbf{r}+\boldsymbol{\xi}-\mathbf{r}'-\boldsymbol{\xi}'|/c$.

Expanding $G_1$ and $G_2$ in a series in powers of $\xi^{\alpha} - \xi'^{\alpha}$ and introducing moments

$$S^{(\alpha_1\alpha_2...\alpha_n)}(\mathbf{r},\mathbf{r}',\mathbf{p},\mathbf{p}',t,t-\tilde{\tau}) =$$
$$= \frac{1}{\Delta^2}\sum_{i=1}^{N}\sum_{k\neq i}^{N}\int_{\Delta(\mathbf{r})}d^3\xi\int_{\Delta(\mathbf{r}')}d^3\xi'(\xi^{\alpha_1}-\xi'^{\alpha_1})(\xi^{\alpha_2}-\xi'^{\alpha_2})...(\xi^{\alpha_n}-\xi'^{\alpha_n})\times \qquad (10)$$
$$\times \delta(\mathbf{r}+\boldsymbol{\xi}-\mathbf{r}_i(t))\delta(\mathbf{p}-\mathbf{p}_i(t))\delta(\mathbf{r}'+\boldsymbol{\xi}'-\mathbf{r}_k(t-\tilde{\tau}))\delta(\mathbf{p}'-\mathbf{p}_k(t-\tilde{\tau})),$$

we obtain

$$I = -\frac{\partial}{\partial \mathbf{p}}\int d^3\mathbf{r}'\int d^3\mathbf{p}'\sum_{n=0}^{\infty}\frac{1}{n!}\left[F^{(1)}_{(\alpha_1\alpha_2...\alpha_n)} + F^{(2)}_{(\alpha_1\alpha_2...\alpha_n)}\frac{\mathbf{p}'}{m}\frac{\partial}{\partial \mathbf{r}'}\right]S^{(\alpha_1\alpha_2...\alpha_n)}(\mathbf{r},\mathbf{r}',\mathbf{p},\mathbf{p}',t,t-\tilde{\tau}).$$

Here

$$F^{(k)}_{(\alpha_1\alpha_2...\alpha_n)}(\mathbf{r}-\mathbf{r}') = \left.\frac{\partial^n G_k}{\partial x^{\alpha_1}\partial x^{\alpha_2}...\partial x^{\alpha_n}}\right|_{\xi^{\alpha}-\xi'^{\alpha}=0}.$$

As a result, the law of change in phase density (6) will take the final form

$$\frac{\partial f(\mathbf{r},\mathbf{p},t)}{\partial t} + \frac{\mathbf{p}}{m}\frac{\partial f(\mathbf{r},\mathbf{p},t)}{\partial \mathbf{r}} =$$
$$= \frac{\partial}{\partial \mathbf{p}}\int d^3\mathbf{r}'\int d^3\mathbf{p}'\sum_{n=0}^{\infty}\frac{1}{n!}\left[F^{(1)}_{(\alpha_1\alpha_2...\alpha_n)} + F^{(2)}_{(\alpha_1\alpha_2...\alpha_n)}\frac{\mathbf{p}'}{m}\frac{\partial}{\partial \mathbf{r}'}\right]S^{(\alpha_1\alpha_2...\alpha_n)}(\mathbf{r},\mathbf{r}',\mathbf{p},\mathbf{p}',t,t-\tilde{\tau}). \qquad (11)$$

This is the main equation of this work, on the basis of which a hierarchy of equations for dynamic distribution functions will be built.

Performing the expansion of the moments in powers of the times of interactions retardation, it is not difficult to verify that after replacing

$$\mathbf{r} \to \mathbf{r},\; t \to -t,\; \mathbf{p} \to -\mathbf{p},$$

corresponding to operation of time reversibility, the left-hand side of (11) does not change, and the change in sign on the right-hand side occurs only for terms of the expansion with odd numbers. Consequently, the resulting equation is noninvariant with respect to time reversal and, as a consequence, describes the irreversible behavior of the system of the particles.

To simplify the further analysis, we carry out expansion on retardation time. Moreover, we take into account that $\tilde{\tau} = |\mathbf{r}+\boldsymbol{\xi}-\mathbf{r}'-\boldsymbol{\xi}'|/c \simeq |\mathbf{r}-\mathbf{r}'|/c = \tau$ and

$$\delta(\mathbf{r}'+\boldsymbol{\xi}'-\mathbf{r}_k(t-\tau)) \simeq \delta(\mathbf{r}'+\boldsymbol{\xi}'-\mathbf{r}_k(t)) + \tau\frac{\mathbf{p}_k(t)}{m}\frac{\partial}{\partial \mathbf{r}'}\delta(\mathbf{r}'+\boldsymbol{\xi}'-\mathbf{r}_k(t)).$$

Then, leaving the terms of the order $\tau$ on the right-hand side of equation (11), the main equation can be represented as

$$\frac{\partial f(\mathbf{r},\mathbf{p},t)}{\partial t} + \frac{\mathbf{p}}{m}\frac{\partial f(\mathbf{r},\mathbf{p},t)}{\partial \mathbf{r}} \simeq$$

$$\simeq \frac{\partial}{\partial \mathbf{p}}\int d^3\mathbf{r}'\int d^3\mathbf{p}'\sum_{n=0}^{\infty}\frac{1}{n!}\left[F_{(\alpha_1\alpha_2...\alpha_n)}^{(1)}\left(1+\frac{|\mathbf{r}-\mathbf{r}'|}{c}\frac{\mathbf{p}'}{m}\frac{\partial}{\partial \mathbf{r}'}\right)+F_{(\alpha_1\alpha_2...\alpha_n)}^{(2)}\frac{\mathbf{p}'}{m}\frac{\partial}{\partial \mathbf{r}'}\right]S^{(\alpha_1\alpha_2...\alpha_n)}(\mathbf{r},\mathbf{r}',\mathbf{p},\mathbf{p}',t). \quad (12)$$

Next, we consider important particular cases of the resulting equation (12).

## 4. First approximation

In the case of force fields that vary slightly on the scales $\sim \sqrt[3]{\Delta}$, only the first term can be left in the sum over $n$ in the first approximation. Then equation (12) takes the form:

$$\frac{\partial f(\mathbf{r},\mathbf{p},t)}{\partial t} + \frac{\mathbf{p}}{m}\frac{\partial f(\mathbf{r},\mathbf{p},t)}{\partial \mathbf{r}} =$$

$$= \frac{\partial}{\partial \mathbf{p}}\int d^3\mathbf{r}'\int d^3\mathbf{p}'\left[\frac{\partial U(\mathbf{r}-\mathbf{r}')}{\partial \mathbf{r}} + \frac{\partial U_{eff}(\mathbf{r}-\mathbf{r}')}{\partial \mathbf{r}}\frac{\mathbf{p}'}{m}\frac{\partial}{\partial \mathbf{r}'}\right]S^{(0)}(\mathbf{r},\mathbf{r}',\mathbf{p},\mathbf{p}',t). \quad (13)$$

Here $U_{eff}(\mathbf{r}-\mathbf{r}') = U(\mathbf{r}-\mathbf{r}')|\mathbf{r}-\mathbf{r}'|/c$ and

$$S^{(0)}(\mathbf{r},\mathbf{r}',\mathbf{p},\mathbf{p}',t) =$$

$$= \frac{1}{\Delta^2}\sum_{i=1}^{N}\sum_{k\neq i}^{N}\int_{\Delta(\mathbf{r})}d^3\xi\int_{\Delta(\mathbf{r}')}d^3\xi'\delta(\mathbf{r}+\xi-\mathbf{r}_i(t))\delta(\mathbf{p}-\mathbf{p}_i(t))\delta(\mathbf{r}'+\xi'-\mathbf{r}_k(t))\delta(\mathbf{p}'-\mathbf{p}_k(t)).$$

is the moment of zero order. Taking into account the Newton's third law for the particles we get $S^{(0)}(\mathbf{r},\mathbf{r}',\mathbf{p},\mathbf{p}',t) = f(\mathbf{r},\mathbf{p},t)f(\mathbf{r}',\mathbf{p}',t)$. This allows us to write (13) in the form

$$\frac{\partial f(\mathbf{r},\mathbf{p},t)}{\partial t} + \frac{\mathbf{p}}{m}\frac{\partial f(\mathbf{r},\mathbf{p},t)}{\partial \mathbf{r}} =$$

$$= \frac{\partial f(\mathbf{r},\mathbf{p},t)}{\partial \mathbf{p}}\int d^3\mathbf{r}'\int d^3\mathbf{p}'\left[\frac{\partial U(\mathbf{r}-\mathbf{r}')}{\partial \mathbf{r}} + \frac{\partial U_{eff}(\mathbf{r}-\mathbf{r}')}{\partial \mathbf{r}}\left(\frac{\mathbf{p}'}{m}\frac{\partial}{\partial \mathbf{r}'}\right)\right]f(\mathbf{r}',\mathbf{p}',t).$$

In view of the equalities of type

$$\int d^3\mathbf{r}'\int d^3\mathbf{p}'\frac{\partial U_{eff}(\mathbf{r}-\mathbf{r}')}{\partial \mathbf{r}}\left(\frac{\mathbf{p}'}{m}\frac{\partial}{\partial \mathbf{r}'}\right)f(\mathbf{r}',\mathbf{p}',t) = -\int d^3\mathbf{r}'\int d^3\mathbf{p}'\frac{\partial U_{eff}(\mathbf{r}-\mathbf{r}')}{\partial \mathbf{r}}\frac{\partial f(\mathbf{r}',\mathbf{p}',t)}{\partial t}$$

kinetic equation (12) for a smoothed phase density takes the form

$$\frac{\partial f(\mathbf{r},\mathbf{p},t)}{\partial t} + \frac{\mathbf{p}}{m}\frac{\partial f(\mathbf{r},\mathbf{p},t)}{\partial \mathbf{r}} =$$
$$= \frac{\partial f(\mathbf{r},\mathbf{p},t)}{\partial \mathbf{p}}\frac{\partial}{\partial \mathbf{r}}\int d^3\mathbf{r}'\int d^3\mathbf{p}' U(\mathbf{r}-\mathbf{r}') f(\mathbf{r}',\mathbf{p}',t) - \quad (14)$$
$$- \frac{\partial f(\mathbf{r},\mathbf{p},t)}{\partial \mathbf{p}}\frac{\partial}{\partial \mathbf{r}}\int d^3\mathbf{r}'\int d^3\mathbf{p}' U_{eff}(\mathbf{r}-\mathbf{r}')\frac{\partial f(\mathbf{r}',\mathbf{p}',t)}{\partial t}.$$

If we neglect the retardation (i.e., at $c \to \infty$), the second term in the right-hand side of equation (14) becomes zero. The equation becomes invariant with respect to time reversal and, as a consequence, no longer describes the irreversible behavior of the particle system.

Equation (14) also is valid for microscopic density, but only if the self-action effect is taken into account [10].

## 5. Second approximation

Holding in the sum over $n$ the first two terms, the right side of the main equation (12) can be represented as

$$I = \frac{\partial f(\mathbf{r},\mathbf{p},t)}{\partial \mathbf{p}}\frac{\partial}{\partial \mathbf{r}}\int d^3\mathbf{r}'\int d^3\mathbf{p}' U(\mathbf{r}-\mathbf{r}') f(\mathbf{r}',\mathbf{p}',t) -$$
$$- \frac{\partial f(\mathbf{r},\mathbf{p},t)}{\partial \mathbf{p}}\frac{\partial}{\partial \mathbf{r}}\int d^3\mathbf{r}'\int d^3\mathbf{p}' U_{eff}(\mathbf{r}-\mathbf{r}')\frac{\partial f(\mathbf{r}',\mathbf{p}',t)}{\partial t} +$$
$$+ \frac{\partial}{\partial \mathbf{p}}\int d^3\mathbf{r}'\int d^3\mathbf{p}' \left[\frac{\partial}{\partial x^\alpha}\left(\frac{\partial U(\mathbf{r}-\mathbf{r}')}{\partial \mathbf{r}}\right) + \left(\frac{\partial}{\partial x^\alpha}\left(\frac{\partial U_{eff}(\mathbf{r}-\mathbf{r}')}{\partial \mathbf{r}}\right)\right) - \right.$$
$$\left. - \frac{\partial U(\mathbf{r}-\mathbf{r}')}{\partial \mathbf{r}}\frac{\partial}{\partial x^\alpha}\frac{|\mathbf{r}-\mathbf{r}'|}{c}\right)\frac{\mathbf{p}'}{m}\frac{\partial}{\partial \mathbf{r}'}\right] S^{(\alpha)}(\mathbf{r},\mathbf{r}',\mathbf{p},\mathbf{p}',t).$$

We write the first-order moment

$$S^{(\alpha)}(\mathbf{r},\mathbf{r}',\mathbf{p},\mathbf{p}',t) = \frac{1}{\Delta^2}\sum_{i=1}^{N}\sum_{k \neq i}^{N}\int_{\Delta(\mathbf{r})} d^3\xi \int_{\Delta(\mathbf{r}')} d^3\xi' (\xi^\alpha - \xi'^\alpha)\delta(\mathbf{r}+\xi-\mathbf{r}_i(t))\times$$
$$\times \delta(\mathbf{p}-\mathbf{p}_i(t))\delta(\mathbf{r}'+\xi'-\mathbf{r}_k(t))\delta(\mathbf{p}'-\mathbf{p}_k(t)).$$

in the form

$$S^{(\alpha)}(\mathbf{r},\mathbf{r}',\mathbf{p},\mathbf{p}',t) = h^\alpha(\mathbf{r},\mathbf{p},t) f(\mathbf{r}',\mathbf{p}',t) - h^\alpha(\mathbf{r}',\mathbf{p}',t) f(\mathbf{r},\mathbf{p},t).$$

Here

$$h^\alpha(\mathbf{r},\mathbf{p},t) = \frac{1}{\Delta}\sum_{k}\int_{\Delta(\mathbf{r})} \xi^\alpha d^3\xi \delta(\mathbf{r}+\xi-\mathbf{r}_k(t))\delta(\mathbf{p}-\mathbf{p}_k(t)) \quad (15)$$

is the vector field of the displacements of particles relative to the center of a volume $\Delta(\mathbf{r})$.

Then, after simple transformations, equation (12) can be represented in the final form

$$\frac{\partial f(\mathbf{r},\mathbf{p},t)}{\partial t}+\frac{\mathbf{p}}{m}\frac{\partial f(\mathbf{r},\mathbf{p},t)}{\partial \mathbf{r}}=$$
$$=\frac{\partial}{\partial \mathbf{p}}\int d^3\mathbf{r}'\int d^3\mathbf{p}' f_2(\mathbf{r},\mathbf{p},\mathbf{r}',\mathbf{p}',t)\frac{\partial U(\mathbf{r}-\mathbf{r}')}{\partial \mathbf{r}}+\Omega[f,\mathbf{h}|\mathbf{r},\mathbf{p},t]. \quad (16)$$

Here $f_2(\mathbf{r},\mathbf{p},\mathbf{r}',\mathbf{p}',t)$ is two-particle distribution function (5), which, in a second approximation, depends on the smoothed phase density $f(\mathbf{r},\mathbf{p},t)$ and the vector field (15) $\mathbf{h}(\mathbf{r},\mathbf{p},t)=h^\alpha(\mathbf{r},\mathbf{p},t)\mathbf{e}_\alpha$:

$$f_2(\mathbf{r},\mathbf{p},\mathbf{r}',\mathbf{p}',t)=f(\mathbf{r},\mathbf{p},t)f(\mathbf{r}',\mathbf{p}',t)+$$
$$+\mathrm{div}_{\mathbf{r}'}\left[\mathbf{h}(\mathbf{r},\mathbf{p},t)f(\mathbf{r}',\mathbf{p}',t)-\mathbf{h}(\mathbf{r}',\mathbf{p}',t)f(\mathbf{r},\mathbf{p},t)\right]. \quad (17)$$

The second term on the right-hand side of equation (16) has the form

$$\Omega[f,\mathbf{h}|\mathbf{r},\mathbf{p},t]=$$
$$=-\frac{\partial}{\partial \mathbf{p}}\int d^3\mathbf{r}'\int d^3\mathbf{p}'\left[f(\mathbf{r},\mathbf{p},t)\frac{\partial \mathbf{h}(\mathbf{r}',\mathbf{p}',t)}{\partial t}-\mathbf{h}(\mathbf{r},\mathbf{p},t)\frac{\partial f(\mathbf{r}',\mathbf{p}',t)}{\partial t}\right]\times$$
$$\times\frac{\mathbf{r}-\mathbf{r}'}{c|\mathbf{r}-\mathbf{r}'|}\frac{\partial U(\mathbf{r}-\mathbf{r}')}{\partial \mathbf{r}}-\frac{\partial}{\partial \mathbf{p}}\int d^3\mathbf{r}'\int d^3\mathbf{p}'\left[f(\mathbf{r},\mathbf{p},t)\frac{\partial f(\mathbf{r}',\mathbf{p}',t)}{\partial t}+\right. \quad (18)$$
$$\left.+\mathrm{div}_{\mathbf{r}'}\left[\mathbf{h}(\mathbf{r},\mathbf{p},t)\frac{\partial f(\mathbf{r}',\mathbf{p}',t)}{\partial t}-f(\mathbf{r},\mathbf{p},t)\frac{\partial \mathbf{h}(\mathbf{r}',\mathbf{p}',t)}{\partial t}\right]\right]\frac{\partial U_{\mathit{eff}}(\mathbf{r}-\mathbf{r}')}{\partial \mathbf{r}}.$$

Due to the term $\Omega$, the dynamics of the system is irreversible. This term corresponds to the influence of the field, through which the interaction between the particles occurs. The influence of the field is taken into account by means of retardation potentials (8). We can say that (18) is connected with a thermostat, the role of which is played not by the external environment, but by the field. There is a continuous exchange of energy between the system of particles and the field, which leads to irreversible evolution.

It should be noted that equation (16) formally has the same form as in the works devoted to the classical theory of kinetic equations [13]. However, the fundamental difference is that the term $\Omega$ in these works is introduced into the theory additionally, from the outside, and corresponds to the interaction with the external environment (thermostat). We also note that the proposed derivation of $\Omega$ is the physical justification of the so-called source term introduced by Zubarev [7], which selects delayed solutions of the kinetic equation.

It is easy to see that in the absence of retardation (i.e., in the case $\Omega=0$), equation (16) is the first equation of the BBGKY hierarchy. In this case, formula (17) gives a recipe for calculating (in a second approximation) the two-particle distribution function and, as a consequence, the correlation functions inherent in statistical physics. For this, it is necessary to find the functions (4) and (15), equations for which are discussed below.

**6. Construction of hierarchy of kinetic equations**

When the next (third) term in the expansion of $n$ is taken into account, tensor quantities of the second rank appear

$$Q^{\alpha\beta}(\mathbf{r},\mathbf{p},t) = \frac{1}{\Delta}\sum_k \int_{\Delta(\mathbf{r})} \xi^\alpha \xi^\beta d^3\xi \delta(\mathbf{r}+\boldsymbol{\xi}-\mathbf{r}_k(t))\delta(\mathbf{p}-\mathbf{p}_k(t)). \qquad (19)$$

These and multipole quantities with higher tensor dimensions are intended for description of the heterogeneity in the distribution of particles. It is possible to continue sequentially accounting the degree of heterogeneity, obtaining a more accurate expression for the two-particle distribution function (5). The reason for the heterogeneity of the particle distribution is due to the inconstancy of the gradient of interparticle fields inside $\Delta(\mathbf{r})$.

Since no probabilistic measures were used in definitions (4), (5), (15), and (19), the origin of the distribution functions has an exclusively deterministic explanation that does not require statistical interpretation.

It follows from (14) that, in the first approximation, the two-particle function corresponds to the so-called mean field approximation $f_2(\mathbf{r},\mathbf{p},\mathbf{r}',\mathbf{p}',t) = f(\mathbf{r},\mathbf{p},t)f(\mathbf{r}',\mathbf{p}',t)$, in which the inhomogeneity in the particle distribution is not taken into account. It is customary to say that in the mean-field approximation, correlations in the arrangement of particles are neglected. From a dynamic point of view, the mean field approximation corresponds to the constancy of the gradient of internal fields within the volumes $\Delta(\mathbf{r})$. In this approximation, kinetic equation (14) is closed.

Since in the second approximation the kinetic equation (16) already contains two unknown functions $f(\mathbf{r},\mathbf{p},t)$ and $\mathbf{h}(\mathbf{r},\mathbf{p},t)$, it is necessary to write another equation to solve the problem. To do this, find the partial time derivative of $\mathbf{h}(\mathbf{r},\mathbf{p},t)$. If we will not take into account tensor quantities of the second rank (as well as in (23)), then the obtained (vector) equation for $\mathbf{h}(\mathbf{r},\mathbf{p},t)$ will contain the functions $f(\mathbf{r},\mathbf{p},t)$ and $\mathbf{h}(\mathbf{r},\mathbf{p},t)$ only. The system becomes close. Having solved it, we can find in this approximation the two-particle distribution function (17).

In the third approximation, the tensor distribution functions (19) will be additionally taken into account. The expression for $f_2(\mathbf{r},\mathbf{p},\mathbf{r}',\mathbf{p}',t)$ will have a more complex form. A closed system will already contain three equations. To obtain them, it is necessary in equation (12) leave in the sum of $n$ the first three terms. Then, differentiating $\mathbf{h}(\mathbf{r},\mathbf{p},t)$ with respect to time, we obtain the second (vector) equation, leaving the distribution functions in it no higher than the second rank (19). The third equation (or rather, six equations for the independent components of the symmetric tensor field $Q^{\alpha\beta}$) is obtained by time differentiation of (19) in the same approximation.

For a more detailed account of heterogeneity in the distribution of particles (i.e., take into account the following approximations), one needs to act in a similar way.

**7. Summary and conclusions**

Let us formulate briefly the main conclusions of the study.

a) The kinetic equation (11) is derived for a smoothed phase density (4), which describes the irreversible dynamics of a system of particles.

b) It is shown that the distribution functions (4), (5), as well as (15) and (19), which determine the macroscopic characteristics of a system of particles, are of a *deterministic origin* and do not require the introduction of probability hypotheses for their justification.

c) It has been established that *irreversibility is due to the influence of a field* (playing the role of a thermostat), the accounting of which is finally expressed in the appearance of an additional term $\Omega$ in the kinetic equation.

d) In our approach, to describe the irreversible behavior of the system, there is no need to introduce additional assumptions and any source term [7] or terms related to them that do not follow from the equations of motion [13].

e) A closure procedure is proposed for a system of integro-differential equations for smoothed distribution functions of different tensor dimensions.

**Acknowledgments**

This work has been supported by the Ministry of Education and Science of Russian Federation within the framework of the project part of the state order (project No. 3.3572.2017/4.6)

Zubkov V.V. grateful to Prof. A.Yu. Zakharov for useful discussions.

**References**

[1] H. J. Kreuzer, Non-Equilibrium Thermodynamics and its Statistical Foundations, Oxford, University Press, 1981.

[2] G.E. Uhlenbeck, G.W. Ford, Lectures in Statistical Mechanics, Providence, Rhode Island, American Mathematical Society, 1963.

[3] G.A. Martynov, Classical Statistical Mechanics, Springer, Netherlands, 1997.

[4] V. V. Yevstafiev, Phase density method: A microscopic description of the gas of neutral particles, Phys Rev E. 64 (2001) 041201. https://doi.org/10.1103/PhysRevE.64.041201

[5] Yu. L. Klimontovich, Statistical Theory of Open Systems: Volume 1: A Unified Approach to Kinetic Description of Processes in Active Systems, Dordrecht, Netherlands, Springer, 1995.

[6] N.N. Bogolyubov, Collection of scientific works. T. V. Nonequilibrium statistical mechanics. Ed. N.M. Plakida, A.D. Sukhanov, Nauka: Moscow, 2005.

[7] D.N.Zubarev, V.G.Morozov, G. Rópke, Statistical mechanics of nonequilibrium processes Vol.1, Wiley-VCH, Akademie Verlag GmbH, Berlin 1996.

[8] N.N. Bogolyubov, Collection of scientific works. T. VI. Equilibrium statistical mechanics. Ed. N.M. Plakida, A.D. Sukhanov Nauka: Moscow, 2006.

[9] M. Kac, Some remarks on the use of probability in classical statistical mechanics, Bull. Acad. Roy. Belg.. 42 (1956) 356-361.

[10] A. Yu. Zakharov, Determinism vs. statistics in classical many-body theory: Dynamical origin of irreversibility, Physica A: Statistical Mechanics and its Applications 473 (2017) 72-76. https://doi.org/10.1016/j.physa.2017.01.005

[11] A. Yu. Zakharov, On physical principles and mathematical mechanisms of the phenomenon of irreversibility, Physica A: Statistical Mechanics and its Applications 525 (2019) 1289-1295. https://doi.org/10.1016/j.physa.2019.04.047

[12] M. A. Drofa, L. S. Kuz'menkov, Theoretical and Mathematical Physics 108 (1996) 848-859. https://doi.org/10.1007/BF02070512

[13] U. M. B. Marconi, S. Melchionna, Kinetic theory of correlated fluids: From dynamic density functional to Lattice Boltzmann methods, J. Chem. Phys. 131 (2009) 014105. https://doi.org/10.1063/1.3166865